\documentclass[twocolumn,preprintnumbers,amsmath,amssymb]{revtex4}
\usepackage{graphicx}
\usepackage{dcolumn}
\usepackage{bm}
\usepackage{times}
\usepackage[colorlinks,citecolor=blue,linkcolor=red]{hyperref}
\usepackage{color}
\usepackage{comment}

\begin{document}

\title{Impact of counter-rotating-wave term on quantum heat transfer and phonon statistics in nonequilibrium qubit-phonon hybrid system}

\author{Chen Wang$^{1,}$}\email{wangchenyifang@gmail.com}
\author{Lu-Qin Wang$^{2}$}
\author{Jie Ren$^{2,}$}\email{Xonics@tongji.edu.cn}
\address{
$^{1}$Department of Physics, Zhejiang Normal University, Jinhua 321004, Zhejiang , P. R. China\\
$^{2}$Center for Phononics and Thermal Energy Science, China-EU Joint Center for Nanophononics, \\
Shanghai Key Laboratory of Special Artificial Microstructure Materials and Technology,  \\
School of Physics Science and Engineering, Tongji University, Shanghai 200092, China
}

\date{\today}

\begin{abstract}
Counter-rotating-wave terms (CRWTs) are traditionally viewed to be crucial in open small quantum systems with strong system-bath dissipation.
Here by exemplifying in a nonequilibrium qubit-phonon hybrid model, we show that CRWTs can play the significant role in  quantum heat transfer even with weak system-bath dissipation.
By using coherent phonon states, we obtain the quantum master equation with heat exchange rates contributed by rotating-wave-terms (RWTs) and CRWTs, respectively.
We find that including only RWTs, steady state heat current and current fluctuations will be significantly suppressed at large temperature bias, whereas they are strongly enhanced by considering CRWTs in addition. Furthermore, for the phonon statistics, the average phonon number and two-phonon correlation are nearly insensitive to strong qubit-phonon hybridization with only RWTs, whereas they will be dramatically cooled down via the cooperative transitions based on CRWTs in addition.
Therefore, CRWTs in quantum heat transfer system should be treated carefully.
\end{abstract}

\maketitle

\section{Introduction}

Understanding and managing nonequilibrium energy transfer at nanoscale is a long-standing problem,
which has attracted great attention with scientific interest and practical importance~\cite{gchen2005book,ydubi2011rmp,nbli2012rmp}.
Theoretically, the microscopic description of the system-bath coupling in open small quantum systems is crucial
to model the quantum heat flow, because the system-bath coupling describes the dissipation of the small quantum system to the external environment~\cite{hpbreuer2007book,uweiss2008book}.
In particular, the strong couplings between the small quantum system and surrounding baths significantly contribute
to the transient and steady state heat transport properties, including
 energy exchange dynamics~\cite{jlsmith2016jcp,akato2016jcp,mcarrega2016prl,jcerrillo2016prb,hmaguire2019prl,wdou2020prb}, quantum thermodynamics and thermal machines~\cite{mesposito2005prx,gkatz2016entropy,wdou2018prb,mpllobet2018prl,kgoyal2019prr,pstrasberg2019prl,arivas2020prl},
and quantum heat transfer~\cite{dsegal2006prb,lnicolin2011prb,lnicolin2011jcp,akato2015jcp,cwang2015sr,cwang2017pra,jjliu2017pre,aqmu2017njp,mbrenes2020prx}.

One of the most representative paradigms of the system-bath dissipation is the spin-boson (qubit-bath) coupling~\cite{ajleggett1987rmp}
\begin{eqnarray}
\hat{V}_{\textrm{SB}}&=&\hat{\sigma}_x\sum_{k}(g_{k}\hat{b}^\dag_{k}+g^*_{k}\hat{b}_k)=\hat{V}_{\textrm{RWT}}+\hat{V}_{\textrm{CRWT}},  \nonumber
\end{eqnarray}
where
\begin{eqnarray}
\hat{V}_{\textrm{RWT}}&=&\sum_{k}(g_{k}\hat{b}^\dag_{k}\hat{\sigma}_-+g^*_{k}\hat{b}_k\hat{\sigma}_+), \nonumber\\
\hat{V}_{\textrm{CRWT}}&=&\sum_{k}(g_{k}\hat{b}^\dag_{k}\hat{\sigma}_++g^*_{k}\hat{b}_k\hat{\sigma}_-),  \nonumber
\end{eqnarray}
with $\hat{\sigma}_x$ the Pauli operator of the central qubit, or say, a two-level system, $g_{k}$ the qubit-bath coupling strength, $\hat{b}^\dag_{k}~(\hat{b}_{k})$ the bosonic creator (annihilator) in the thermal bath. This system-bath dissipation form has been widely used for the heat transport in quantum phononics. The rotating-wave-terms (RWTs) $\hat{V}_{\textrm{RWT}}$ describe energy conserved processes that can occur sequentially in an incoherent picture, while the counter-rotating-wave-terms (CRWTs)
$\hat{V}_{\textrm{CRWT}}$ depict two energy non-conserved processes that can only occur coherently.

It is well believed that in the weak qubit-bath dissipation limit, $\hat{V}_{\textrm{CRWT}}$ is negligible and $\hat{V}_{\textrm{RWT}}$ dominates the dissipation process.
Only when the dissipation strength increases, $\hat{V}_{\textrm{CRWT}}$ is necessarily included to properly characterize the heat exchange~\cite{uweiss2008book,mcarrega2016prl}.
Specifically, in quantum dissipative dynamics, $\hat{V}_{\textrm{CRWT}}$ is found to significantly enhance anti-Zeno signal~\cite{hzheng2008prl,zhli2009pra,xfcao2010pra,qai2010pra},
quantum correlation~\cite{agd2010prl,jma2012pra,cwang2013njp} and
spontaneous emission~\cite{yli2012pra,yli2013pra,syang2013pra} at strong system-bath coupling.
At steady state, the non-canonical statistical properties of the small quantum system is apparently exhibited~\cite{cklee2012pre,dzxu2014pre,jsmith2014pra,dzxu2016fp}, and the heat current is optimally strengthened~\cite{dsegal2006prb,cwang2015sr,jjliu2017pre}.
Recently, due to the fast development of circuit-QED with high quality microwave resonators, the spin-boson coupling is reduced to the qubit-photon hybridization
$\hat{V}_{\textrm{SB}}=\hat{\sigma}_x(g\hat{b}^\dag+g^*\hat{b})$ with a single mode of bosons (photons)~\cite{pfd2019rmp,afk2019nrp}.
The importance of CRWTs $\hat{V}_{\textrm{CRWT}}=(g\hat{b}^\dag\hat{\sigma}_++g^*\hat{b}\hat{\sigma}_-)$ in photonic hybrid quantum systems on energy level structure has been experimental demonstrated
~\cite{tn2010np,fy2017np}.
Moreover, from the aspect of photon statistics,  the qubit-photon interaction ($\hat{V}_{\textrm{SB}}$) will affect
the two-photon correlation function in dressed picture at equilibrium to exhibit novel features~\cite{ridolfo2012prl,ridolfo2013prl,rstassi2013prl}.


For steady state heat transfer with two baths at nonequilibrium setup,
the nonequilibrium spin-boson (qubit-bath) dissipation
$\hat{V}_{\textrm{NSB}}=\hat{\sigma}_x\sum_{k;v=L,R}(g_{k,v}\hat{b}^\dag_{k,v}+g^*_{k,v}\hat{b}_{k,v})$
is considered generic to establish the thermodynamic bias between phononic baths L and R~\cite{dsegal2005prl,dsegal2008prl}.
Specifically, in the weak qubit-bath interaction limit, the heat flow shows sequential process, which is dominated by RWTs.
In sharp contrast, the heat current and fluctuation exhibit cooperative energy exchange processes between two phononic baths
at strong qubit-bath coupling~\cite{dsegal2006prb,lnicolin2011jcp,lnicolin2011prb,cwang2015sr}, where CRWTs play the leading role.
While under the adiabatic modulation of two bath temperatures, the finite heat pump based on the Redfield approximation is exhibited with weak qubit-bath dissipation~\cite{jren2010prl}, which is irrelevant with the dissipation strength.
However, in strong dissipation limit the heat pump can be observed only for the biased qubit~\cite{tchen2012prb,cwang2017pra}.
Moreover, the exploration of such distinction originating from RWTs and CRTs has been extended to the quantum thermal transistor from weak to strong spin-boson couplings~\cite{kjoulain2016prl,bqguo2018pre,bqguo2019pre,jydu2019pre,cwang2018pra,hliu2019pre,hfyang2020jpb,cwang2020cpb}.

We investigated the steady state heat current 
in nonequilibrium qubit-phonon hybrid model
at weak qubit-phonon hybridization regime~\cite{cwang2020arxiv}. Here, we propose that at strong qubit-phonon hybridization inside the central system, CRWTs of the (even weak) qubit-bath dissipation may novelly contribute to the transport behaviors of hybrid quantum systems.
The main points of this work have three manifolds:
(i) Based on the quantum master equation encoded with coherent phonon states,
we obtain microscopic pictures of heat exchange processes, which are separately contributed by RWTs and CRWTs.
(ii) By including full counting statistics with weak system-bath interactions,
the heat current and fluctuations are strongly suppressed at large temperature bias
with the qubit-bath dissipation considering only RWTs, whereas the heat current fluctuations are greatly enhanced by considering CRWTs in addition.
 This clearly demonstrates the novel role of CRWTs on steady state heat transfer.
(iii) The steady state phonon number and two-phonon correlation function with RWTs are nearly insensitive to the strong qubit-phonon hybridization,
whereas they are optimally cooled down via the cooperative transition path contributed by both RWTs and CRWTs.

This paper is organized as follows:
In Sec. II, we describe the nonequilibrium qubit-phonon hybrid model in part A, derive the quantum master equation in part B,
and obtain the expression of heat current fluctuations by including full counting statistics in part C.
In Sec. III, we investigate steady state heat transfer, which are characterized by heat current, noise power and skewness.
In Sec. IV, we analyze the steady state phonon number and two-phonon correlation function  of phonon statistics.
Finally, we give a conclusion in Sec. V.

\section{Model and method}

\subsection{Qubit-phonon hybrid system}
The nonequilibrium qubit-phonon hybrid system, which is composed by two-level qubits interacting with a single vibrational mode, each individually coupled to a bosonic thermal bath, is described as~($\hbar=1$)
\begin{eqnarray}~\label{h0}
\hat{H}=\hat{H}_s+\sum_{u=qu,ph}(\hat{H}_u+\hat{V}_u).
\end{eqnarray}
Specifically, the hybrid system is expressed as
\begin{eqnarray}~\label{hs0}
\hat{H}_s=\varepsilon\hat{J}_z+\omega_0\hat{a}^\dag\hat{a}+{\lambda}(\hat{a}^\dag+\hat{a})\hat{J}_z,
\end{eqnarray}
where the collective qubit operators are $\hat{J}_{\alpha}=\frac{1}{2}\sum^{N}_{i=1}\hat{\sigma}^i_{\alpha}~(\alpha=x,y,z)$
with $N$ the qubit number and $\hat{\sigma}^i_{\alpha}$ is the Pauli operator of the $i$th qubit,
$\varepsilon$ is the energy splitting of the qubits,
$\hat{a}^{\dag}~(\hat{a})$ creates(annihilates) one phonon with the frequency $\omega_0$,
and $\lambda$ is the hybridization strength between the qubits and the phononic field.
The $u$th thermal bath is described as
$\hat{H}_u=\sum_{k}\omega_k\hat{b}^\dag_{k,u}\hat{b}_{k,u}$,
where $\hat{b}^\dag_{k,u}~(\hat{b}_{k,u})$ creates (annihilates) one boson in the $u$th thermal bath with the momentum $k$ and frequency $\omega_{k,u}$.
The interaction between the phononic mode and the $ph$th bath is given by
\begin{eqnarray}~\label{vph0}
\hat{V}_{ph}=\sum_{k}(g_{k,ph}\hat{b}^{\dag}_{k,c}+g^{*}_{k,ph}\hat{b}_{k,c})(\hat{a}^{\dag}+\hat{a}),
\end{eqnarray}
with $g_{k,ph}$ the coupling strength.
While for the interaction  between the qubits and the $qu$th thermal bath, the interaction can be generally expressed as
\begin{eqnarray}~\label{vqu0}
\hat{V}_{qu}=\sum_k(g_{k,qu}\hat{b}^{\dag}_{k,qu}\hat{S}+g^{*}_{k,q}\hat{b}_{k,q}\hat{S}^\dag),
\end{eqnarray}
with $g_{k,qu}$ the coupling strength between the qubits and the $qu$th thermal bath.
For the qubit operator $\hat{S}$, if we analyze $\hat{V}_{qu}$ within the rotating wave approximation (RWA), it is specified as
$\hat{S}=\hat{J}_-$.
Hence,  under the influence of rotating-wave terms (RWTs) [i.e., $\hat{V}^{\textrm{RWT}}_{qu}=\sum_k(g_{k,qu}\hat{b}^{\dag}_{k,qu}\hat{J}_-+g^{*}_{k,q}\hat{b}_{k,qu}\hat{J}_+)$], the particle number of the whole system is conserved as the exchange processes occur between the qubits and the $qu$th bath.
However,  when we consider the full interaction between the qubits and the $qu$th bath $\hat{V}^{\textrm{full}}_{qu}$, the operator becomes $\hat{S}=2\hat{J}_x$,
which both includes the RWTs and the counter-rotating-wave terms (CRWTs) [i.e., $\hat{V}^{\textrm{CRT}}_{qu}=\sum_k(g_{k,qu}\hat{b}^{\dag}_{k,qu}\hat{J}_++g^{*}_{k,q}\hat{b}_{k,qu}\hat{J}_-)$].
Under the effect of CRWTs, both the system and corresponding thermal bath will be excited (annihilated) simultaneously, which apparently breaks the particle number (energy) conservation. We would like to point out although the angular momentum conservation is not explicitly considered at present, it is implicitly contained in the spin-boson dissipation, where the raising and lowering of spin angular momentum  is compensated by the coupled bosons due to their intrinsic spin~\cite{phononspin1,phononspin2}.
Moreover, it needs to stress that the spin-boson and qubit-bath dissipation(coupling) denote the spin coupled to bosonic thermal bath, which is composed by continuous boson modes.
While the qubit-phonon hybridization means the interaction between the spin and the single mode phononic resonator.

In this work, we mainly analyze the novel role of CRWTs on the steady state behaviors of the hybrid quantum system at Eq.~(\ref{h0}).
For the the qubit-phonon hybrid system $\hat{H}_s$ at Eq.~(\ref{hs0}), the eigensolution can be exactly solved as
$\hat{H}_s|\phi^k_m{\rangle}=E^k_{m}|\phi^k_m{\rangle}$.
The eigenvalue is
\begin{eqnarray}~\label{emk}
E^k_{m}=\varepsilon{m}+\omega_0k-\lambda^2m^2/\omega_0,
\end{eqnarray}
and the eigenvector is
\begin{eqnarray}
|\phi^k_m{\rangle}=|j,m{\rangle}{\otimes}[\frac{(\hat{a}^\dag+g_m)^k}{\sqrt{k!}}|0{\rangle}_{m}],
\end{eqnarray}
where the phonon excitation number is $k=0,1,2,...$,  the angular momentum state is $\hat{J}_z|j,m{\rangle}=m|j,m{\rangle}~(m=-j,-j+1,...,j)$ with $j=N/2$, the displacement coefficient is $g_m={\lambda}m/\omega_0$,
the coherent phonon state is $|0{\rangle}_{m}=e^{-g_m\hat{a}^\dag-g^2_m/2}|0{\rangle}_a$,
and the bare vacuum state is $\hat{a}|0{\rangle}_a=0$.


It should be noted that though the nonequilibrium qubit-phonon hybrid system is theoretically investigated in the present paper,
it has experimental correspondences.
Specifically, the hybrid quantum system can be specified by the system composed by the nanomechanical resonator and single quantum dot~\cite{pstadler2014prl},
where the resonator and quantum dot interact with the bosonic thermal and magnon reservoir~\cite{jren2013prb1,jren2013prb2}, respectively.
The CRWTs at Eq.~(\ref{vqu0}) can be realized by the interfacial interaction with non-spin-conservation~\cite{mmatsuo2018prl}.
Moreover, it could also be realized by the circuit-QED setup~\cite{mmajland2020prb},
where one josephson junction could be longitudinally coupled to a \textrm{LC} resonator~\cite{pmbil2015prb,sricher2016prb}.
While for the spin-boson model, under the reaction coordinate mapping approach it can be mapped to another type of qubit-phonon hybrid model~\cite{agarg1985jcp,mthoss2001jcp,jismith2014pra,jlsmith2016jcp,gschaller2016pre,pstrasberg2018prb,hmaguire2019prl}.
From the aspect of inverse design, the analysis of quantum heat transfer in  nonequilibrium qubit-phonon hybrid systems could fertilize theoretical interact and practical application of the spin-boson model.

\begin{figure}[tbp]
\begin{center}
\includegraphics[scale=0.4]{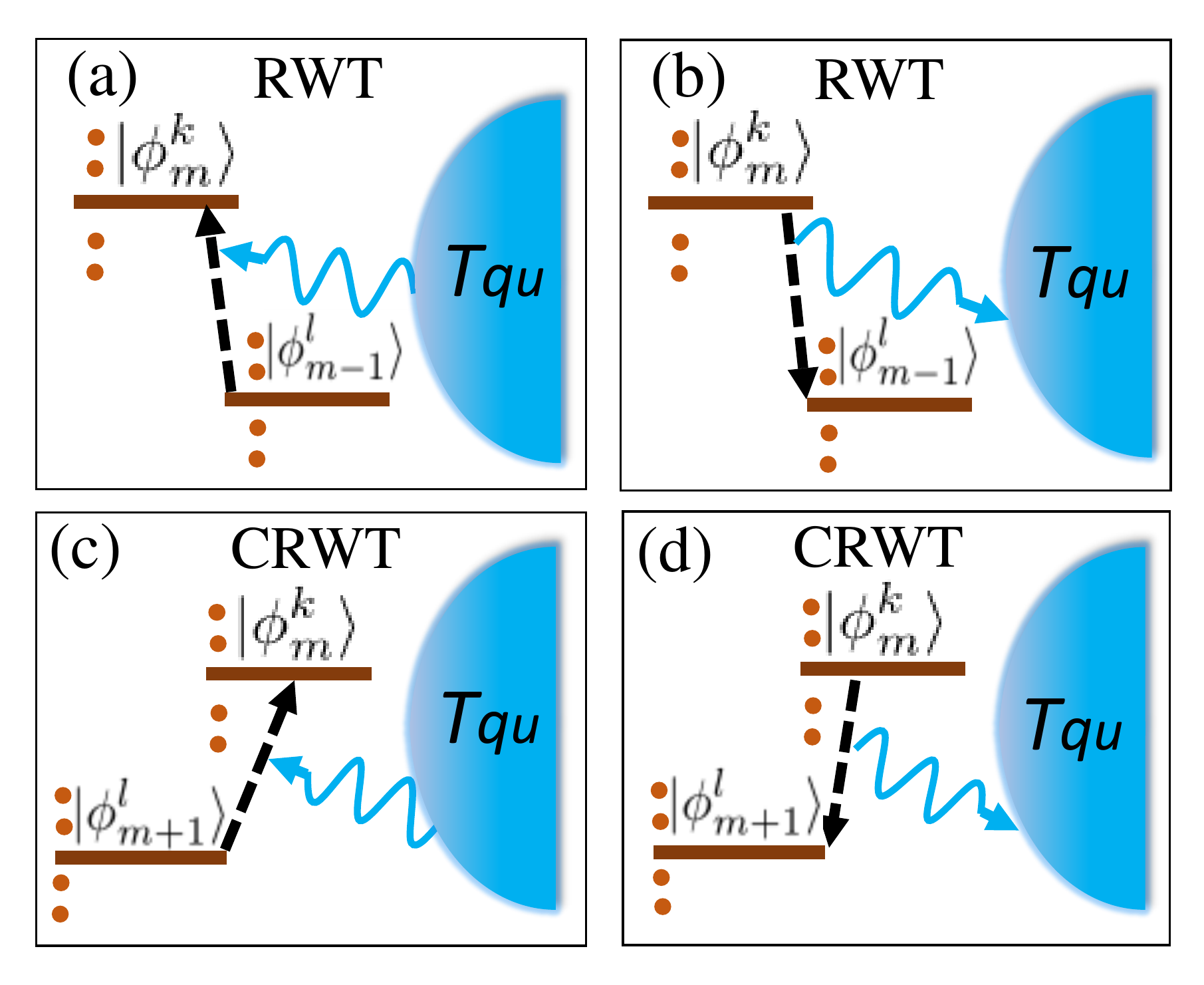}
\end{center}
\caption{(Color online)
Microscopic heat transfer processes assisted by the $qu$th thermal bath.
Under the effect of RWTs:
(a) the excitation process from the coherent phonon state $|\phi^l_{m-1}{\rangle}$ to $|\phi^k_m{\rangle}$
by absorbing one boson from the $qu$th bath with the energy $E^{m-1,l}_{m,k}$ characterized by the rate
$\Gamma^{+}_{qu,\textrm{RWT}}(\phi^l_{m-1}|\phi^k_m)$;
(b) the relaxation process from the coherent phonon state $|\phi^k_m{\rangle}$ to $|\phi^l_{m-1}{\rangle}$
by releasing one boson into the $qu$th bath with the energy $E^{m-1,l}_{m,k}$ characterized by the rate
$\Gamma^{-}_{qu,\textrm{RWT}}(\phi^l_{m-1}|\phi^k_m)$.
Under the effect of CRWTs:
(c) the excitation process from the coherent phonon state  $|\phi^l_{m+1}{\rangle}$ to $|\phi^k_{m}{\rangle}$
by absorbing one boson from the $qu$th bath with the energy $E^{m+1,l}_{m,k}$ characterized by the rate
$\Gamma^{+}_{qu,\textrm{CRWT}}(\phi^l_{m+1}|\phi^k_{m})$;
(d) the relaxation process from  $|\phi^k_{m}{\rangle}$ to $|\phi^l_{m+1}{\rangle}$
by releasing one boson into the $qu$th bath with the energy $E^{m+1,l}_{m,k}$ characterized by the rate
$\Gamma^{-}_{qu,\textrm{CRWT}}(\phi^l_{m+1}|\phi^k_{m})$.
}~\label{fig1}
\end{figure}

\subsection{Quantum master equation}
Considering weak system-bath interactions, we separately perturb $\hat{V}_{ph}$ and $\hat{V}_{qu}$ to obtain the quantum master equation.
Under the Born-Markov approximation, the total density operator is decomposed as
$\hat{\rho}_{tot}(t){\approx}\hat{\rho}_{s}(t){\otimes}\hat{\rho}_b$,
where $\hat{\rho}_{s}(t)$ is the reduced density operator of qubit-phonon hybrid system,
and $\hat{\rho}_b=\exp(-\sum_{u=ph,qu}\hat{H}^u_b/k_BT_u)/\mathcal{Z}$ is the equilibrium density operator of thermal baths,
with $k_B$ the Boltzmann constant, $T_u$ the temperature of $u$th thermal bath
and  $\mathcal{Z}=\textrm{Tr}\{\exp(-\sum_{u=ph,qu}\hat{H}^u_b/k_BT_u)\}$ the partition function.
In this paper, we set $k_B=1$ for convenience.
Then, by tracing over the degrees of freedom of thermal baths, the generalized master equation is obtained at Eq.~(\ref{gqme1}) in Appendix.

From the generalized quantum master equation,
it is known that for transient dynamics the populations are generally coupled to the off-diagonal terms~\cite{asettineri2018pra}.
However, after long time evolution, it is numerically checked in a wide parameter regime that the off-diagonal terms become negligible.
Hence, the generalized quantum master equation is reduced to the dressed master equation as
\begin{eqnarray}~\label{dme1}
\frac{d\hat{\rho}_s(t)}{dt}&=&\mathcal{\hat{L}}_0\hat{\rho}_s(t)+
\sum\{
\Gamma^+_u(E^{m^{\prime}k^{\prime}}_{mk})\mathcal{\hat{L}}_+(|\phi^k_m{\rangle}{\langle}\phi^{k^\prime}_{m^\prime}|)\hat{\rho}_s(t)\nonumber\\
&&+\Gamma^-_u(E^{m^{\prime}k^{\prime}}_{mk})\mathcal{\hat{L}}_-(|\phi^{k^\prime}_{m^\prime}{\rangle}{\langle}\phi^k_m|)\hat{\rho}_s(t)\},
\end{eqnarray}
where the dissipators are given by
\begin{eqnarray}
&&\mathcal{\hat{L}}_0\hat{\rho}_s(t)=-i[\hat{H}_s,\hat{\rho}_s(t)]\nonumber\\
&&-\frac{1}{2}\sum
\Gamma^+_u(E^{m^{\prime}k^{\prime}}_{mk})(|\phi^{k^\prime}_{m^\prime}{\rangle}{\langle}\phi^{k^\prime}_{m^\prime}|\hat{\rho}_s+
\hat{\rho}_s|\phi^{k^\prime}_{m^\prime}{\rangle}{\langle}\phi^{k^\prime}_{m^\prime}|)\nonumber\\
&&-\frac{1}{2}\sum\Gamma^-_u(E^{m^{\prime}k^{\prime}}_{mk})
(|\phi^{k}_{m}{\rangle}{\langle}\phi^{k}_{m}|\hat{\rho}_s+
\hat{\rho}_s|\phi^{k}_{m}{\rangle}{\langle}\phi^{k}_{m}|),\\
&&\mathcal{\hat{L}}_+(|\phi^k_m{\rangle}{\langle}\phi^{k^\prime}_{m^\prime}|)\hat{\rho}_s=
|\phi^k_m{\rangle}{\langle}\phi^{k^\prime}_{m^\prime}|\hat{\rho}_s(t)|\phi^{k^\prime}_{m^\prime}{\rangle}{\langle}\phi^k_m|,\\
&&\mathcal{\hat{L}}_-(|\phi^k_m{\rangle}{\langle}\phi^{k^\prime}_{m^\prime}|)\hat{\rho}_s=
|\phi^{k^\prime}_{m^\prime}{\rangle}{\langle}\phi^k_m|\hat{\rho}_s(t)|\phi^k_m{\rangle}{\langle}\phi^{k^\prime}_{m^\prime}|.
\end{eqnarray}
The nonzero rates assisted by the $ph$th bath are
\begin{eqnarray}~\label{gph0}
\Gamma^{\pm}_{ph}(\phi^{k-1}_m|\phi^{k}_m)={\pm}k\gamma_{ph}(E^{m,k-1}_{m,k})n_{ph}({\pm}E^{m,k-1}_{m,k}),
\end{eqnarray}
with the energy gap $E^{m,k-1}_{m,k}=E_{m,k}-E_{m,k-1}$.
$\Gamma^+_{ph}(\phi^{k-1}_m|\phi^{k}_m)~[\Gamma^-_{ph}(\phi^{k-1}_m|\phi^{k}_m)]$ describes the phonon excitation(relaxation) process from the coherent phonon state
$|\phi^{k-1}_m{\rangle}~(|\phi^{k}_m{\rangle})$ to $|\phi^{k}_m{\rangle}~(|\phi^{k-1}_m{\rangle})$
by absorbing(releasing) one boson with the energy $E^{m,k-1}_{m,k}$ from(into) the $ph$th thermal bath,
with the qubits state unchanged.
For the qubit-bath interaction under rotating-wave approximation, the rates assisted by the $qu$th bath are given by
\begin{eqnarray}
\Gamma^{\pm}_{qu,\textrm{RWT}}(\phi^{l}_{m-1}|\phi^{k}_m)&&={\pm}
\theta(E^{m-1,l}_{m,k})(j^+_{m-1})^2D^2_{k,l}(\frac{\lambda}{\omega_0})\nonumber\\
&&{\times}\gamma_{qu}(E^{m-1,l}_{m,k})n_{qu}({\pm}E^{m-1,l}_{m,k}),
\end{eqnarray}
where the herald step function is $\theta(\omega{\ge}0)=1$ and $\theta(\omega{<}0)=0$, the angular momentum factor is $j^{\pm}_{m}=\sqrt{j(j+1)-m(m{\pm}1)}$, the coherent phonon state overlap coefficient is
\begin{eqnarray}
D_{k,l}(x)=e^{-x^2/2}\sum^{\min[k,l]}_{n=0}\frac{(-1)^n\sqrt{k!l!}x^{k+l-2n}}{(k-n)!(l-n)!n!},
\end{eqnarray}
and the energy gap is $E^{m-1,l}_{m,k}=E_{m,k}-E_{m-1,l}$.
The rate $\Gamma^+_{qu,\textrm{RWT}}(\phi^{l}_{m-1}|\phi^{k}_m)~[\Gamma^-_{qu,\textrm{RWT}}(\phi^{l}_{m-1}|\phi^{k}_m)]$
characterizes the microscopic transfer process from the coherent phonon state
$|\phi^{l}_{m-1}{\rangle}~(|\phi^{k}_m{\rangle})$ to $|\phi^{k}_m{\rangle}~(|\phi^{l}_{m-1}{\rangle})$
by exchange $|l-k|$ phonon number and the energy $E^{m-1,l}_{m,k}$ involved with the $qu$th thermal bath in Fig.~\ref{fig1}(a) [Fig.~\ref{fig1}(b)],
which is simultaneously bounded by the unidirectional transition of the qubits state from $|j,m-1{\rangle}~(|j,m{\rangle})$ to $|j,m{\rangle}~(|j,m-1{\rangle})$.
While for the full qubit-bath interaction both including RWTs and CRWTs, the transition rate assisted by the $qu$th bath is expressed as
$\Gamma^{\pm}_{qu}(\phi^{l}_{n}|\phi^{k}_m)=\Gamma^{\pm}_{qu,\textrm{RWT}}(\phi^{l}_{m-1}|\phi^{k}_m)\delta_{n,m-1}+
\Gamma^{\pm}_{qu,\textrm{CRWT}}(\phi^{l}_{m+1}|\phi^{k}_m)\delta_{n,m+1}$,
where the rates contributed by CRWTs are given by
\begin{eqnarray}
\Gamma^{\pm}_{qu,\textrm{CRWT}}(\phi^{l}_{m+1}|\phi^{k}_m)&&={\pm}\theta(E^{m+1,l}_{m,k})
(j^-_{m+1})^2D^2_{k,l}(\frac{\lambda}{\omega_0})\nonumber\\
&&{\times}\gamma_{qu}(E^{m+1,l}_{m,k})n_{qu}({\mp}E^{m+1,l}_{m,k}),
\end{eqnarray}
with the positive energy gap  $E^{m+1,l}_{m,k}=E_{m,k}-E_{m+1,l}$.
For $\Gamma^+_{qu}(\phi^{l}_{n}|\phi^{k}_m)~[\Gamma^-_{qu}(\phi^{l}_{n}|\phi^{k}_m)]$, it is found that besides the transfer processes mastered by $\Gamma^+_{qu,\textrm{RWT}}(\phi^l_{m-1}|\phi^k_m)~[\Gamma^-_{qu,\textrm{RWT}}(\phi^l_{m-1}|\phi^k_m)]$,
the rate  also describes the another distinct transition from the state $|\phi^{l}_{m+1}{\rangle}~(|\phi^{k}_{m}{\rangle})$
to $|\phi^{k}_{m}{\rangle}~(|\phi^{l}_{m+1}{\rangle})$ in Fig.~\ref{fig1}(c)[Fig.~\ref{fig1}(d)],
which is characterized by the rate component $\Gamma^+_{qu,\textrm{CRWT}}(\phi^{l}_{m+1}|\phi^{k}_m)~[\Gamma^-_{qu,\textrm{CRWT}}(\phi^{l}_{m+1}|\phi^{k}_m)]$.
It is noted that this process is accompanied by the angular momentum transition from $|j,m+1{\rangle}~(|j,m{\rangle})$ to $|j,m{\rangle}~(|j,m+1{\rangle})$.
This additional transfer processes contributed by CRWTs will significantly affect the steady state features of the hybrid quantum system, e.g., nonequilibrium heat transfer and phonon statistics, even with weak qubit-bath dissipation.



\subsection{Quantum master equation combined with FCS}

We focus on the steady state heat transfer, which includes the steady state current and current fluctuations.
We add the counting parameter to count the energy flow into the $ph$th thermal bath based on the dressed master equation at Eq.~(\ref{dme1}),
where the off-diagonal elements become negligible.
Specifically, we first introduce the
generalized density operator $\hat{\rho}(t,Q_t)$,
where $Q_t$ is the transferred energy into the $ph$th bath during the time interval $t$.
Accordingly, the master equation can be described as
\begin{eqnarray}~\label{dmechi}
\frac{d\hat{\rho}_s(t,Q_t)}{dt}&=&\hat{\mathcal{L}}_0\hat{\rho}_s(t,Q_t)\nonumber\\
&&+\sum\{
\Gamma^+_u(E^{m^{\prime}k^{\prime}}_{mk})
\mathcal{\hat{L}}_+(|\phi^k_m{\rangle}{\langle}\phi^{k^\prime}_{m^\prime}|)\hat{\rho}^+_s(t,Q_t)\nonumber\\
&&+\Gamma^-_u(E^{m^{\prime}k^{\prime}}_{mk})
\mathcal{\hat{L}}_-(|\phi^{k^\prime}_{m^\prime}{\rangle}{\langle}\phi^k_m|)\hat{\rho}^-_s(t,Q_t)\}.
\end{eqnarray}
with $\hat{\rho}^{\pm}_s(t,Q_t)=\hat{\rho}_s(t,Q_t{\mp}E^{m^{\prime}k^{\prime}}_{mk})$.
The probability of counting the energy $Q_t$ after the time $t$ is given by
$P(t,Q_t)=\textrm{Tr}\{\hat{\rho}_s(t,Q_t)\}$.
Then, we apply a Fourier transformation in the energy space by including the counting parameter $\chi$
via $\hat{\rho}_s(t,\chi)=\sum_{Q_t}\hat{\rho}_s(t,Q_t)e^{i{\chi}Q_t}$.
This leads to the modified dressed master equation
\begin{eqnarray}~\label{dmechi2}
\frac{d\hat{\rho}_s(t,\chi)}{dt}&=&\hat{\mathcal{L}}_0\hat{\rho}_s(t,\chi)\nonumber\\
&&+\sum\{
\Gamma^+_u(E^{m^{\prime}k^{\prime}}_{mk})
\mathcal{\hat{L}}^\chi_+(|\phi^k_m{\rangle}{\langle}\phi^{k^\prime}_{m^\prime}|)\hat{\rho}_s(t,\chi)\nonumber\\
&&+\Gamma^-_u(E^{m^{\prime}k^{\prime}}_{mk})
\mathcal{\hat{L}}^\chi_-(|\phi^{k^\prime}_{m^\prime}{\rangle}{\langle}\phi^k_m|)\hat{\rho}_s(t,\chi)\}.
\end{eqnarray}
where
$\mathcal{\hat{L}}^\chi_{+}(|\phi^k_m{\rangle}{\langle}\phi^{k^\prime}_{m^\prime}|)=
e^{({-}i{\chi}E^{m^{\prime}k^{\prime}}_{mk}\delta_{u,ph})}
\mathcal{\hat{L}}_{+}(|\phi^k_m{\rangle}{\langle}\phi^{k^\prime}_{m^\prime}|)$,
$\mathcal{\hat{L}}^\chi_{p-}(|\phi^{k^\prime}_{m^\prime}{\rangle}{\langle}\phi^k_m|)=
e^{(i{\chi}E^{m^{\prime}k^{\prime}}_{mk}\delta_{u,ph})}
\mathcal{\hat{L}}_{-}(|\phi^{k^\prime}_{m^\prime}{\rangle}{\langle}\phi^k_m|)$,
$\delta_{u,ph}=1$ for $u=ph$, and $\delta_{u,ph}=0$ for $u=qu$.p
Hence, the steady state cumulant generating function is obtained as
$\mathcal{Z}(\chi)=\lim_{t{\rightarrow}\infty}\frac{1}{t}\ln[\textrm{Tr}\{\hat{\rho}_s(t,\chi)\}]$.
Consequently, the $n$th cumulant of steady state heat flow is given by
\begin{eqnarray}
J^{(n)}=\frac{{\partial}^n\mathcal{Z}(\chi)}{{\partial}(i\chi)^n}\Big{|}_{\chi=0}.
\end{eqnarray}
In particular, the steady state heat current is
$J=\frac{{\partial}\mathcal{Z}(\chi)}{{\partial}(i\chi)}\Big{|}_{\chi=0}$,
the noise power is
$J^{(2)}=\frac{{\partial}^2\mathcal{Z}(\chi)}{{\partial}(i\chi)^2}\Big{|}_{\chi=0}$,
and the skewness is
$J^{(3)}=\frac{{\partial}^3\mathcal{Z}(\chi)}{{\partial}(i\chi)^3}\Big{|}_{\chi=0}$.

\begin{figure}[tbp]
\begin{center}
\vspace{-5.0cm}
\includegraphics[scale=0.5]{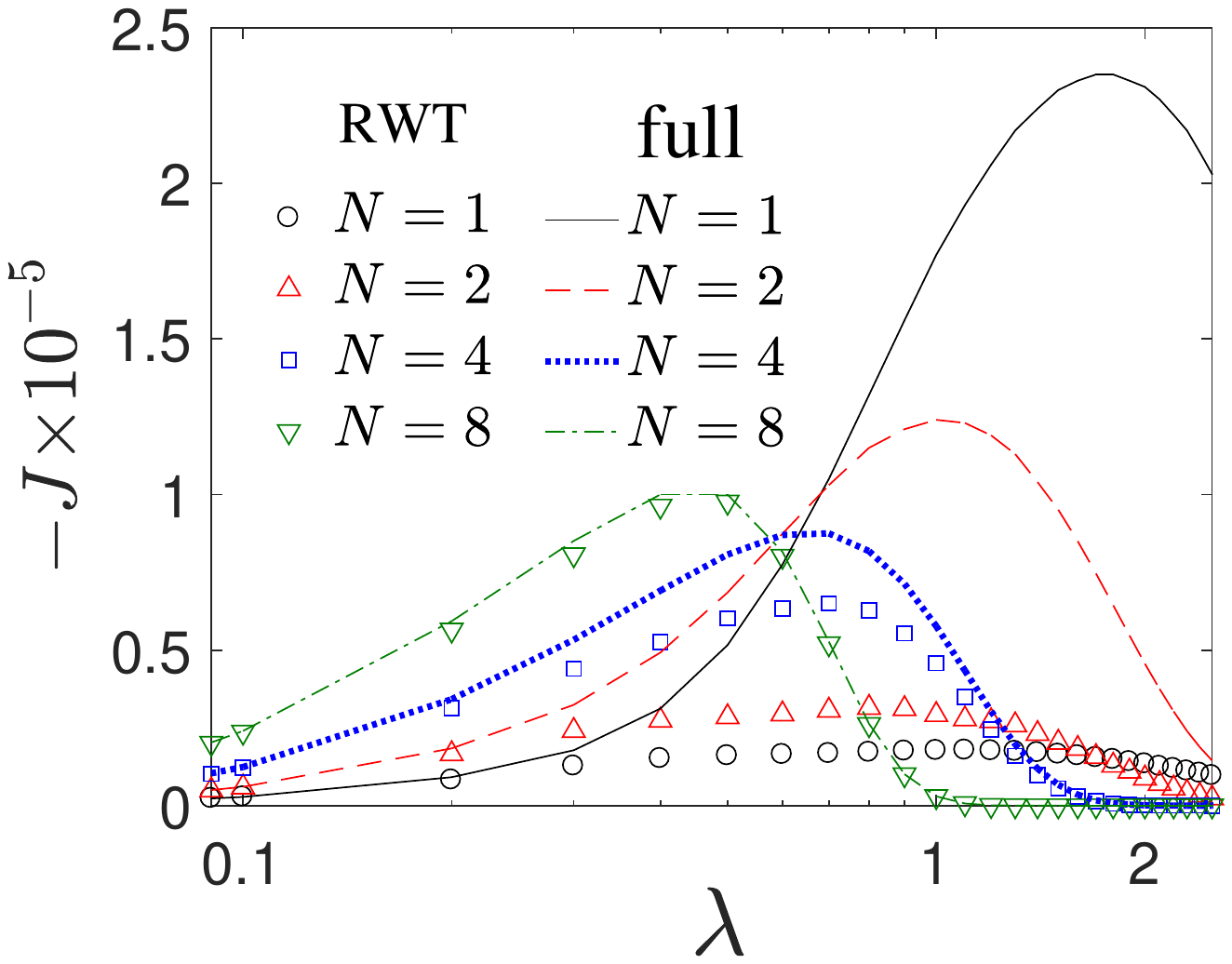}
\vspace{-5.0cm}
\end{center}
\caption{(Color online)
Steady state heat current out of the $ph$th thermal bath as a function of the qubit-phonon hybridization strength under the effects of
only RWTs and full qubit-bath interaction(including both RWTs and CRWTs), respectively.
The other system parameters are given by
$\varepsilon=1$, $\omega_0=1$, $\alpha_{ph}=\alpha_{qu}=0.005$, $T_{ph}=1.5$, $T_{qu}=0.5$,
and $\omega_c=10$.
}~\label{fig11}
\end{figure}

We plot the heat current at resonance in Fig.~\ref{fig11} as one typical instance to analyze the effect of the finite qubit number $N$ on
 steady state behaviors.
It is found that in the small qubit number limit(e.g., $N=1,2$) the heat current with the full interaction between the qubits and $qu$th bath is dramatically enhanced over a wide qubit-phonon coupling regime, compared to the counterpart affected by only RWTs.
This demonstrates the nontrivial contribution from CRWTs.
However, as the qubit number becomes large(e.g., $N=8$),  the currents from two different types of qubit-bath interactions(i.e., RWTs and full interaction) become nearly identical, which implies that RWTs dominate the behavior of heat current.
Therefore, we select $N=1$ in the following to manifest the novel role of CRWTs.

\section{Quantum heat transfer}

\begin{figure}[tbp]
\begin{center}
\includegraphics[scale=0.3]{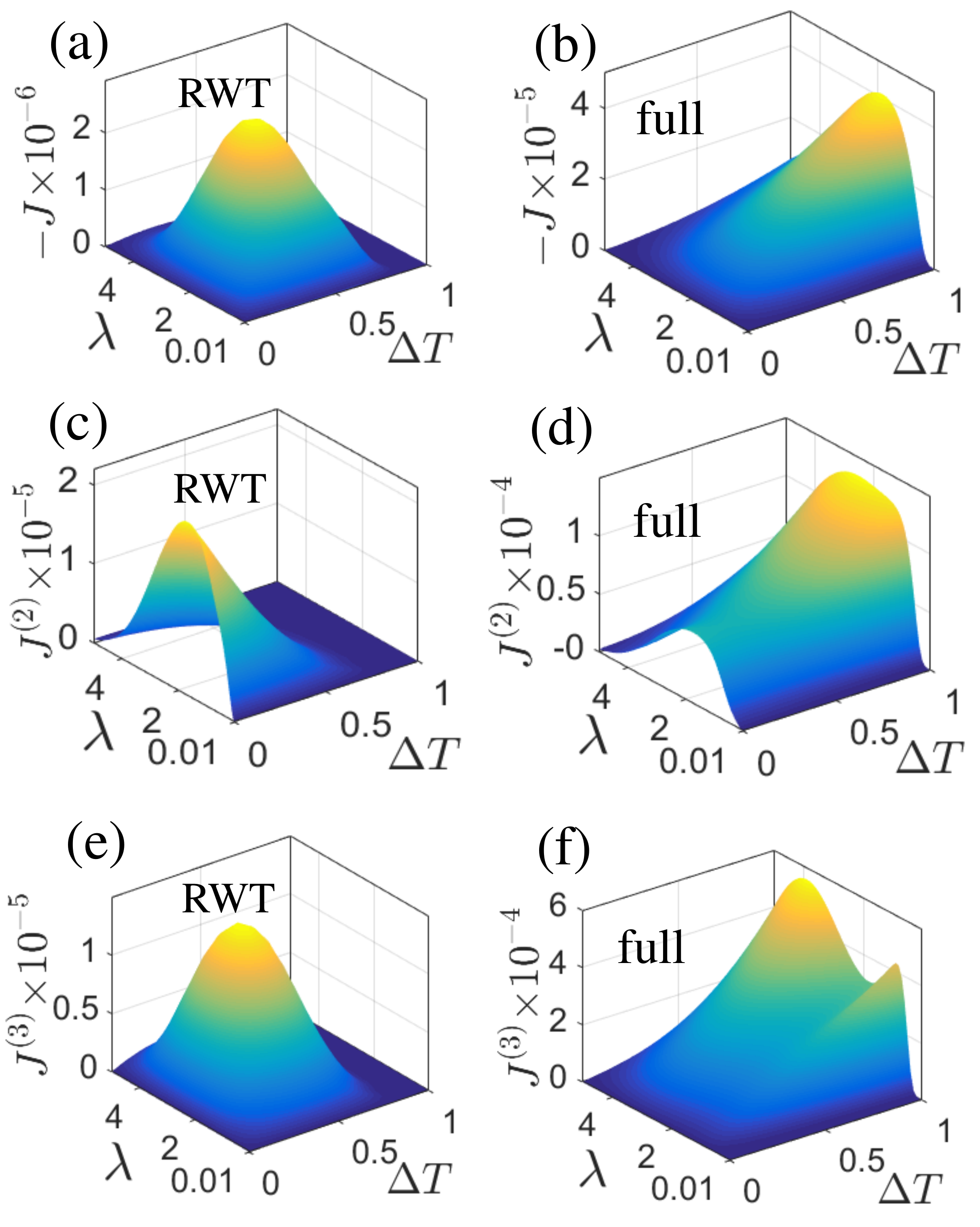}
\end{center}
\caption{(Color online)
(a) and (b) Steady state heat current($-J$),
(c) and (d) noise power$[J^{(2)}]$,
(e) and (f) skewness$[J^{(3)}]$ by modulating both the temperature bias $\Delta{T}$ and qubit-phonon hybridization strength $\lambda$
under the effects of RWTs and CRWTs, with $T_{ph}=T_0+{\Delta}T$, $T_{qu}=T_0-{\Delta}T$, and $T_0=1$.
The other system parameters are given by
$\varepsilon=1$, $\omega_0=1$, $N=1$, $\alpha_{ph}=\alpha_{qu}=0.005$,
and $\omega_c=10$.
}~\label{fig2}
\end{figure}



In this section, we investigate heat current, noise power and skewness at steady state, which are the representative characteristics of heat current fluctuations.
In one previous work, we mainly investigated the steady state heat current and the effect of negative differential thermal conductance(NDTC) with weak qubit-phonon hybridization~\cite{cwang2020arxiv}, where the role of CRWTs was not explicitly explored.
Here, we focus on the comparison of heat current fluctuations owning RWTs at strong qubit-phonon hybridization  with the counterpart including both RWTs and CRWTs.
This will unravel the novel contribution of CRWTs to steady state heat transfer.

We first study the steady state heat current out of the $ph$th thermal bath~($-J$) in Fig.~\ref{fig2}.
Under the influence of  RWTs, Fig.~\ref{fig2}(a) exhibits a globally optimal peak at finite temperature bias (i.e., ${\Delta}T{\approx}0.5$) and strong qubit-phonon hybridization strength (i.e., $\lambda{\approx}2$).
This clearly demonstrates that the effect of NDTC can also be observed with strong qubit-phonon hybridization.
In sharp contrast, by including both  RWTs and CRWTs we find that the heat current shows monontonic enhancement by increasing the temperature bias in the strong hybridization regime, shown in Fig.~\ref{fig2}(b).
Moreover, we  unravel such distinction from the aspect of heat current fluctuations.
Specifically, for the noise power and skewness with only RWTs in Fig.~\ref{fig2}(c) and Fig.~\ref{fig2}(e), they are significantly suppressed at large temperature bias(e.g., $T_{ph}{\approx}2$ and $T_{qu}{\approx}0$),
Whereas the counterparts show dramatic enhancement by including CRWTs in Fig.~\ref{fig2}(d) and Fig.~\ref{fig2}(f).
Hence, we conclude that CRWTs nontrivially enhance steady state heat transfer with strong qubit-phonon hybridization, particularly in large temperature bias regime.

Next, we try to explore the underlying mechanism for the difference of heat current fluctuations.
Under the effect of only RWTs at large temperature bias limit,
the excitation transition quantified by the rate $\Gamma^+_{qu,\textrm{RWT}}(\phi^l_{-\frac{1}{2}}|\phi^k_{\frac{1}{2}})$ naturally vanishes, due to $n_{qu}(E^{-\frac{1}{2},l}_{\frac{1}{2},k})=0$.
 Consequently, the relaxation transition from the coherent phonon state with higher angular momentum state($|\frac{1}{2},\frac{1}{2}{\rangle}$) to the coherent phonon state with lower counterpart($|\frac{1}{2},-\frac{1}{2}{\rangle}$) dominates the heat exchange processes between the dressed qubit and the $qu$th bath, which is characterized by $\Gamma^-_{qu,\textrm{RWT}}(\phi^l_{-\frac{1}{2}}|\phi^k_{\frac{1}{2}})$.
This lead to the population depletion of the coherent phonon states associated with $P_{\frac{1}{2},k}$.
And the nonzero steady state populations become $P_{-\frac{1}{2},k}=e^{-k\omega_0/(k_BT_{ph})}[1-e^{-\omega_0/(k_BT_{ph})}]$,
which are fully thermalized by the $ph$th bath.
Finally, tphis prevents the hybrid system from establishing the thermodynamic bias to drivep steady state heat current and fluctuations,
resulting in the persistent suppression of the current fluctuations.


While for  steady state heat transfer including both  RWTs and CRWTs at large temperature bias,
though the spin flip-up transition accompanied by the energy excitation in Fig.~\ref{fig1}(a) driven by the RWTs is suppressed,
such spin-flip transition can be alternatively realized by releasing one boson into the $qu$th bath under the effect of CRWTs
in Fig.~\ref{fig1}(d).
Hence, under the cooperative contributions of the $qu$th bath assisted transfer processes in Fig.~\ref{fig1}(b) and Fig.~\ref{fig1}(d)
and the processes accompanied by the $ph$th bath characterized by the rates $\Gamma^{\pm}_{ph}(\phi^{k-1}_m|\phi^k_m)$ at Eq.~(\ref{gph0}),
the steady state populations $P_{\frac{1}{2},k}$ can be dramatically excited by increasing the temperature $T_{ph}$.
Simultaneously, a completely thermodynamic cycle can be restored.
This could explain the enhancement of the heat current in Fig.~\ref{fig2}(b) and current fluctuations  in Fig.~\ref{fig2}(d) and Fig.~\ref{fig2}(f).



\begin{figure}[tbp]
\begin{center}
\includegraphics[scale=0.35]{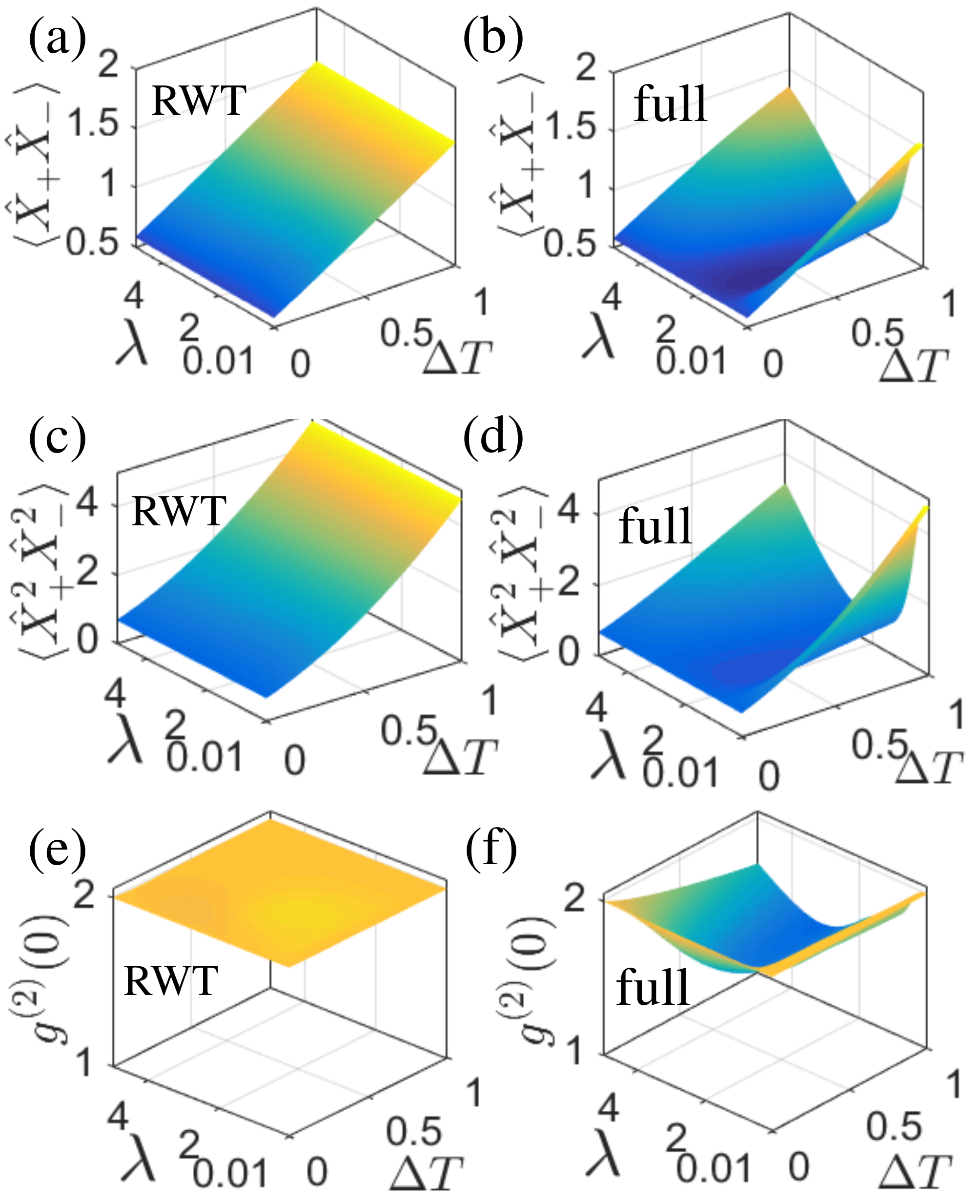}
\end{center}
\caption{(Color online)
Representative quantities of phonon statistics at steady state by tuning both the qubit-phonon hybridization strength $\lambda$
and temperature bias ${\Delta}T$,
which include only RWTs and full interactions for weak qubit-bath coupling, respectively.
(a) and (b) show the expectation values of the phonon number ${\langle}\hat{X}_+\hat{X}_-{\rangle}$;
(c) and (d) are the two-phonon correlation terms ${\langle}\hat{X}^2_+\hat{X}^2_-{\rangle}$;
(e) and (f) are zero-time delay two-phonon correlation functions $g^{(2)}(0)$.
The other system parameters are the same as in Fig.~\ref{fig2}.
}~\label{fig3}
\end{figure}

\section{Two-Phonon Statistics}


In the quantum theory of optical coherence, the zero-time delay two-photon correlation function was initially defined by R. J. Glauber as~\cite{rglauber1963pr}
\begin{eqnarray}
g^{(2)}_a(0)=\frac{{\langle}\hat{a}^\dag\hat{a}^\dag\hat{a}\hat{a}{\rangle}}{{\langle}\hat{a}^\dag\hat{a}{\rangle}^2},
\end{eqnarray}p
where $\hat{a}^\dag~(\hat{a})$ creates(annihilates) one photon in the cavity, and $\langle{\hat{A}}\rangle$ denotes the expectation value of the operator $\hat{A}$.
$g^{(2)}_a(0)$ is traditionally applied to study the statistical features of photons.
Specifically, the bunching effect of the photon-photon correlation with super-Poisson distribution is characterized as
$g^{(2)}_a(0)>1$,
whereas $g^{(2)}_a(0)<1$ as the two-photon statistics becomes antibunching with sub-Poisson distribution.
Moreover, the correlation function $g^{(2)}_a(0)=2$ for the thermal state~\cite{hjc2008book}.

However, it was later proposed that such definition of $g^{(2)}_a(0)$ may only be properly adopted to study the photon statistics with weak light-matter hybridization~\cite{ridolfo2012prl,ridolfo2013prl,rstassi2013prl,lgarziano2013pra,dpagel2015pra,qbin2019pra,hgxu2020jpb}.
As the light-matter interaction becomes strong, the two-photon correlation function should be modified in the dressed picture of the hybrid quantum system~\cite{ridolfo2012prl}
\begin{eqnarray}~\label{g20}
g^{(2)}(0)=\frac{{\langle}\hat{X}^2_+\hat{X}^2_-{\rangle}}{{\langle}\hat{X}_+\hat{X}_-{\rangle}^2},
\end{eqnarray}
where the transition projector is $\hat{X}_-=-i\sum_{k>j}\Delta_{kj}X_{jk}|{\phi}_j{\rangle}{\langle}\phi_k|$,
$\hat{X}_+=(\hat{X}_-)^\dag$, the energy gap is $\Delta_{kj}=E_k-E_j$,
 the transition coefficient is $X_{jk}={\langle}\phi_j|(\hat{a}^\dag+\hat{a})|\phi_k{\rangle}$,
and $|\phi_k{\rangle}$ the eigenstate of the hybrid system.
Physically, $\hat{X}_-~(\hat{X}_+)$ describes the relaxing(exciting) transfer process from the eigenstate $|\phi_k{\rangle}~(|\phi_l{\rangle})$
to the eigenstate $|\phi_l{\rangle}~(|\phi_k{\rangle})$, which is bounded by $\Delta_{kj}>0$.
For the qubit-phonon hybrid system the transition projector under the coherent phonon state basis is generally expressed as
\begin{eqnarray}
\hat{X}_-=-i\omega_0\sum_{m,k}\sqrt{k}|\phi^{k-1}_m{\rangle}{\langle}\phi^k_m|,
\end{eqnarray}
and  the definitions of $g^{(2)}_a(0)$ and $g^{(2)}(0)$  are different.
However, in the weak qubit-phonon hybridization limit(i.e., $\lambda/\omega_0{\approx}0$),
the transition operator is simplified to $\hat{X}_-=-i\hat{a}$,
and the one and two phonon correlation terms are specified as
\begin{eqnarray}~\label{tpc1}
{\langle}\hat{X}_+\hat{X}_-{\rangle}=\omega^2_0n_{ph}(\omega_0),~{\langle}\hat{X}^2_+\hat{X}^2_-{\rangle}=2\omega^4_0n^2_{ph}(\omega_0),
\end{eqnarray}
which leads to the zero-time delay two-phonon correlation function is obtained as $g^{(2)}(0)=2$.

Here, we adopt the definition of the correlation function at Eq.~(\ref{g20}) to investigate the two-phonon statistics at steady state in Fig.~\ref{fig3}.
We first study the phonon statistics at thermal equilibrium(i.e., $T_{ph}=T_{qu}=T_0$).
The populations are given by
$P_{m,k}=e^{(-E_{m,k}/k_BT_0)}/[\sum_{m,k}e^{(-E_{m,k}/k_BT_0)}]~(m={\pm}1/2;k=0,1,2,...)$, which is valid both with and without CRWTs.
Then, the average phonon number is ${\langle}\hat{X}_+\hat{X}_-{\rangle}=\omega^2_0n_{ph}(\omega_0)$, two-phonon correlation term is
${\langle}\hat{X}^2_+\hat{X}^2_-{\rangle}=2\omega^4_0n^2_{ph}(\omega_0)$ and two-phonon correlation function is $g^{(2)}(0)=2$.
It should be noted that this result is analytically obtained for arbitrary qubit-phonon hybridization strength, which is distinct from the counterpart at Eq.~(\ref{tpc1}) at weak qubit-phonon hybridization limit.
Hence, the CRWTs show negligible contribution to the phonon statistics at thermal pequilibrium.

Next, we investigate the phonon statistics at the finite thermodynamic bias.
For the average phonon number with RWTs in Fig.~\ref{fig3}(a), it is generally insensitive to the qubit-phonon hybridization strengthp, and becomes most significant in the bias limit $T_{ph}{\approx}2$ and $T_{qu}{\approx}0$.
The reason is that in the low temperature regime of $T_{qu}$,
the $qu$th bath main assists the unidirectional transition from $|\frac{1}{2},\frac{1}{2}{\rangle}$ branch of coherent phonon states
to the $|\frac{1}{2},-\frac{1}{2}{\rangle}$ branch of coherent phonon states.
The mechanism is quite similar for the two-phonon correlation term ${\langle}\hat{X}^2_+\hat{X}^2_-{\rangle}$, which is exhibited in Fig.~\ref{fig3}(c).
On the contrary, for ${\langle}\hat{X}_+\hat{X}_-{\rangle}$ with full qubit-bath interaction,
it is interesting to find that by increasing the qubit-phonon hybridization strength to strong coupling(e.g., $\lambda=2$), the average phonon number is dramatically suppressed, shown in Fig.~\ref{fig3}(b).
Compared to the transitions only with RWTs, the full qubit-bath interaction include additional transition channels to significantly cool down the phonon field, i.e.,
$|\phi^{l}_{-1/2}{\rangle}{\rightarrow}|\phi^{k}_{1/2}{\rangle}{\rightarrow}|\phi^{l^\prime}_{-1/2}{\rangle}$
with the energy restriction $E^{l}_{-1/2}>E^k_{1/2}>E^{l^\prime}_{-1/2}$ and phonon excitation number bias $(l-l^\prime){\ge}1$.
Moreover, the existence of such collective transition paths greatly decreases the magnitude of the two-phonon term
${\langle}\hat{X}^2_+\hat{X}^2_-{\rangle}$ accordingly.
This mainly results in $g^{(2)}(0)$ apparently lower than $2$.
Therefore, we conclude that  CRWTs generate additional novel energy transfer paths to suppress both the average phonon number and two-phonon correlation function.

\section{Conclusion}
To summarize, we have studied steady state statistics of the nonequilibrium qubit-phonon hybrid system by applying quantum master equation under the coherent phonon state basis.
At steady state, the natural vanish of off-diagonal terms of the reduced hybrid system density matrix simplifies the generalized master equation to the dressed mater equation.
For steady state heat transfer, we have adopted the full counting statistics to compare the heat current and current fluctuations both under the effects of
only RWTs and full interaction including both RWTs and CRWTs.

We first investigated the heat current at strong qubit-phonon hybridization.
It has been found that the current always shows the behavior of NDTC under the effect of RWTs.
In particular at the large temperature bias limit,
the unidirectional  transition from the coherent phonon states in spin-up branch to the counterparts in spin-down branch
prevents the hybrid system from establishing a thermodynamic cycle, shown in Fig.~\ref{fig1}(b).
Such suppression mechanism persists also for current fluctuations, e.g., noise power and skewness.
On the contrary, the heat current and current fluctuations contributed by both  RWTs and CRWTs exhibit monotonic increase by increasing the temperature bias.
The novel transition controlled by CRWTs in Fig.~\ref{fig1}(d) restores the thermodynamic cycle.

Moreover, we have analyzed the average phonon number and two-phonon correlation function  at strong qubit-phonon hybridization and finite temperature bias.
It has been shown that under the effect of RWTs  the phonon is mainly thermally distributed in the coherent phonon states with spin-down branch.
The average phonon number is approximately given by ${\langle}\hat{X}_+\hat{X}_-{\rangle}=\omega^2_0n_{ph}(\omega_0)$,
and two-phonon correlation function is $g^{(2)}{\approx}2$,
which are both nearly insensitive to the qubit-phonon hybridization strength.
While including CRWTs, we have discovered that the average phonon number and the two-phonon correlation function are significantly cooled down in the optimally strong hybridization regime,
due to the collective energy down transition
controlled by the processes in Fig.~\ref{pfig1}(b) and Fig.~\ref{fig1}(d).

\section{Acknowledgement}
W.C. is supported by the National Natural Science Foundation of China under Grant No. 11704093
and the Opening Project of Shanghai Key Laboratory of Special Artificial Microstructure Materials and Technology.
W.L.Q. and R.J. acknowledge the support by the National Natural Science Foundation of China (No. 11775159, No. 11935010), the Natural Science Foundation of Shanghai (No. 18ZR1442800 and No. 18JC1410900).

\appendix
\section{Generalized quantum master equation}
\begin{widetext}
By individually perturbing the phonon-bath and qubit-phonon interactions at Eq.~(\ref{vph0}) and Eq.~(\ref{vqu0}),
we obtain the generalized master equation under the Born-Markov approximation as
\begin{eqnarray}~\label{gqme1}
\frac{d\hat{\rho}_s(t)}{dt}&=&-i[\hat{H}_s,\hat{\rho}_s(t)]
+\frac{1}{2}\sum_{u=ph,qu;\omega^\prime>0}\{
\kappa^+_u(\omega^\prime)[\hat{S}^\dag_u\hat{\rho}_s(t)\hat{S}_{u,<}(-\omega^\prime)+H.c.]
+\kappa^-_u(\omega^\prime)[\hat{S}_u\hat{\rho}_s(t)\hat{S}^\dag_{u,<}(-\omega^\prime)+H.c.]\nonumber\\
&&-\kappa^+_u(\omega^\prime)[\hat{S}_u\hat{S}^\dag_{u,<}(-\omega^\prime)\hat{\rho}_s(t)+H.c.]
-\kappa^-_u(\omega^\prime)[\hat{S}^\dag_u\hat{S}_{u,<}(-\omega^\prime)\hat{\rho}_s(t)+H.c.]
\}
\end{eqnarray}
where the operators are $\hat{S}_{ph}=\hat{a}$, $\hat{S}_{qu}=\hat{S}$,
the components are obtained as $\hat{S}_u(-\tau)=
\sum_{\omega>0}[\hat{S}_{u,>}(\omega)e^{-i\omega\tau}+\hat{S}_{u,<}(-\omega)e^{i\omega\tau}]
+\sum_{\omega=0}\hat{S}_{u,0}$, and
the transition rates between two coherent phonon states are
$\kappa^+_u(\omega)=\gamma_u(\omega)n_u(\omega)$,
$\kappa^-_u(\omega)=\gamma_u(\omega)[1+n_u(\omega)]$,
with the Bose-Einstein distribution function
$n_u(\omega)=1/[\exp(\omega/k_BT_u)-1]$.
\end{widetext}


\end{document}